\documentclass[12pt]{iopart}
\usepackage[dvips]{graphicx}
\usepackage{amssymb}
\usepackage{amsfonts}
\newcommand {\dr}{{\mathrm d}\mathbf{r}}

\newcommand {\rr}{\mathbf{r}}

\newcommand {\dd}{{\mathrm d}}

\newcommand {\GG}{\mathbf{G}}

\begin{document}

\title[DDFT: phase separation in a cavity and the influence of symmetry]
{Dynamical density functional theory:
phase separation in a cavity and the influence of symmetry}

\author{A J Archer} 
\address{H H Wills Physics Laboratory, University of Bristol,
Bristol BS8 1TL, UK}
\date{\today}

\begin{abstract}
Consider a fluid composed of two species of particles, where the interparticle
pair potentials $u_{11} = u_{22} \neq u_{12}$. On confining an
equal number of particles from each species in a cavity, one finds
that the average one body density profiles of each species are
constrained to be exactly the same due to the symmetry, when both external
cavity potentials are the same.
For a binary fluid of Brownian particles interacting via repulsive
Gaussian pair potentials that exhibits phase separation, we study the dynamics
of the fluid one body density profiles on breaking the symmetry of the external
potentials, using the dynamical density functional theory of Marconi and
Tarazona [{\it J. Chem. Phys.}, {\bf 110}, 8032 (1999)]. On breaking the
symmetry we see that the fluid one body density profiles can then show the phase
separation that is present.
\end{abstract}


In equilibrium density functional theory (DFT) \cite{Evans92}, the key quantity
is the Helmholtz free energy functional. Given this functional, one can
calculate the (ensemble) average one body density profile(s) for a fluid of
particles subject to any given external
potential. The form of the Helmholtz free energy functional depends, of course,
upon the particular interaction potentials between the particles in the fluid.
However, this functional is in general unknown, except in a few cases for
1--dimensional fluids \cite{Evans92}. None the less, in the last 30 years, a
number
of rather accurate {\em approximate} Helmholtz free energy functionals have been
developed for a wide class of fluids -- see Ref.\ \cite{Evans92} and references
therein. When attempting to construct a theory for the dynamics of the
(ensemble) average density profile of a fluid confined in a time dependent
external potential, it is very appealing, given this large body of work, to
base a theory for the dynamics upon the equilibrium Helmholtz free
energy functional. In other words, if one is wanting to study the dynamics of a
fluid, is it possible to just apply the relevant equilibrium Helmholtz
free energy functional in the dynamical theory?

Recently, Marconi and Tarazona derived a dynamical density functional theory
(DDFT) \cite{Marconi:TarazonaJCP1999,Marconi:TarazonaJPCM2000} for fluids of
Brownian particles. Their theory is based upon the equilibrium fluid Helmholtz
free energy functional in a rather simple way and therefore has
attracted much interest. The theory is not exact; it
involves approximating the two body correlations in the non-equilibrium fluid
by those of an equilibrium fluid with the {\em same} one body density profile
\cite{Marconi:TarazonaJCP1999,Marconi:TarazonaJPCM2000,Archer7}.
However, for all the cases
where comparison with Brownian dynamics (BD) simulations has been made, the
theory has proved to be remarkably accurate \cite{Marconi:TarazonaJCP1999,
joe:christos, flor1, Archer11, RexetalPRE2005}. It appears that as long as one
has an accurate approximation for the equilibrium Helmholtz free energy
functional, then the DDFT gives a good account of the dynamics for Brownian
fluids.

\begin{figure}[t]
\begin{center}
\begin{minipage}[t]{5.1cm}
\includegraphics[width=5cm]{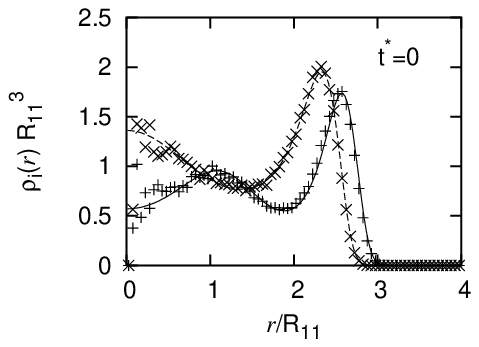}
\end{minipage}
\begin{minipage}[t]{5.1cm}
\includegraphics[width=5cm]{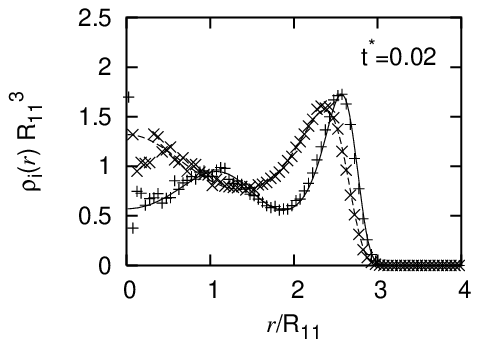}
\end{minipage}
\begin{minipage}[t]{5.1cm}
\includegraphics[width=5cm]{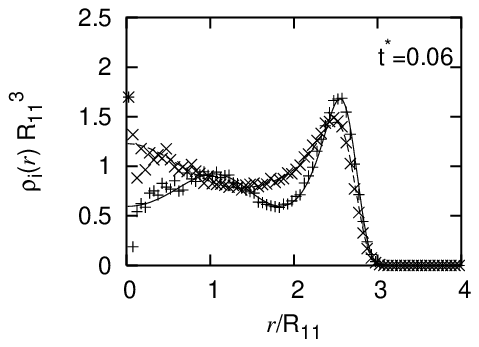}
\end{minipage}
\begin{minipage}[t]{5.1cm}
\includegraphics[width=5cm]{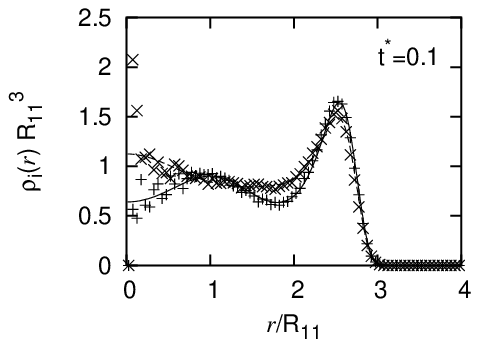}
\end{minipage}
\begin{minipage}[t]{5.1cm}
\includegraphics[width=5cm]{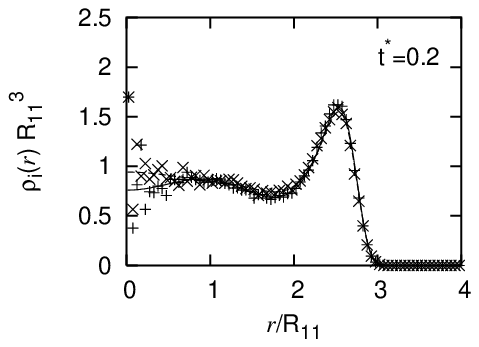}
\end{minipage}
\begin{minipage}[t]{5.1cm}
\includegraphics[width=5cm]{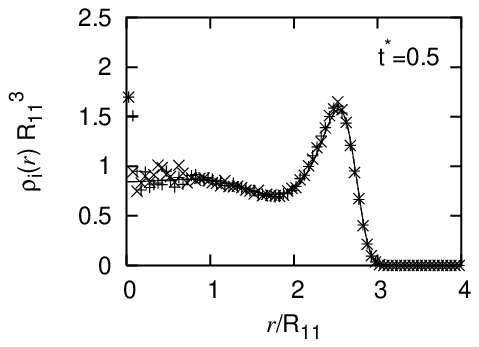}
\end{minipage}
\end{center}
\caption{Fluid A density profiles $\rho_i(r,t)$ (solid line: DDFT results
for species 1, dashed line: species 2;
symbols are BD results, ($+$) for species 1, ($\times$)
for species 2) for $N_1=N_2=100$ particles, which is
initially ($t<0$) at equilibrium confined in (asymmetric)
external potentials with $E_1=10k_BT$,
$E_2=20k_BT$ and ${\cal R}=3R_{11}$. At $t=0$, $E_2$ is suddenly reduced to
$E_2=10k_BT$, giving symmetrical confinement. The profiles are
plotted for various $t^*=k_BT \Gamma R_{11}^2 t$. This model fluid exhibits
microphase-separation: the initial configuration has an `onion' structure, but
on enforcing the symmetry in the external potentials, any signature of this
structure in the one body density profiles disappears as time progresses.}
\label{fig:1}
\end{figure}

\begin{figure}[t]
\begin{center}
\begin{minipage}[t]{5.1cm}
\includegraphics[width=5cm]{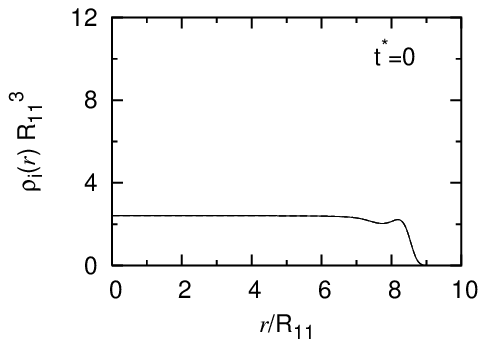}
\end{minipage}
\begin{minipage}[t]{5.1cm}
\includegraphics[width=5cm]{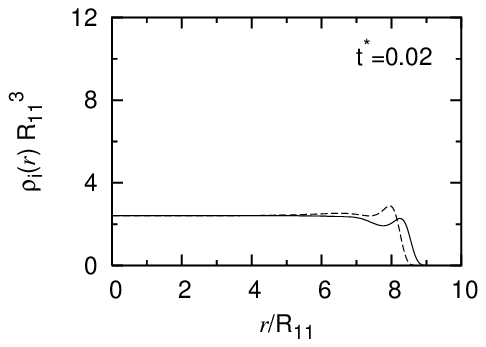}
\end{minipage}
\begin{minipage}[t]{5.1cm}
\includegraphics[width=5cm]{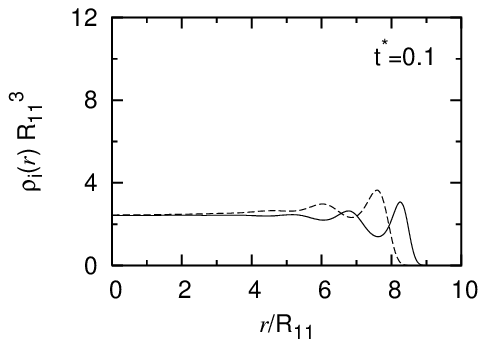}
\end{minipage}
\begin{minipage}[t]{5.1cm}
\includegraphics[width=5cm]{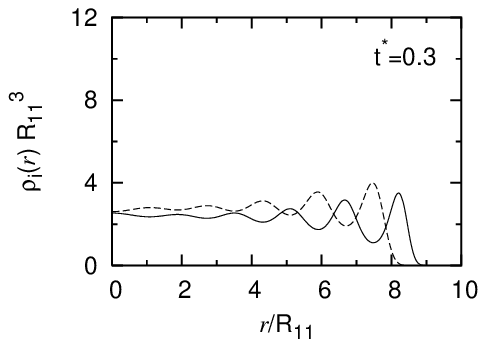}
\end{minipage}
\begin{minipage}[t]{5.1cm}
\includegraphics[width=5cm]{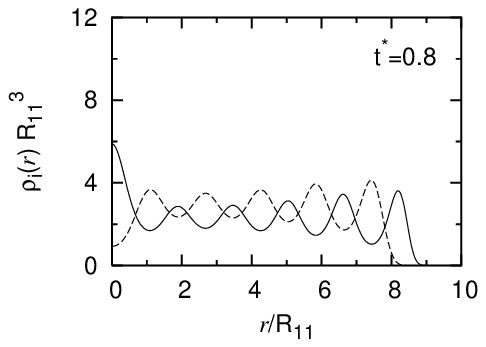}
\end{minipage}
\begin{minipage}[t]{5.1cm}
\includegraphics[width=5cm]{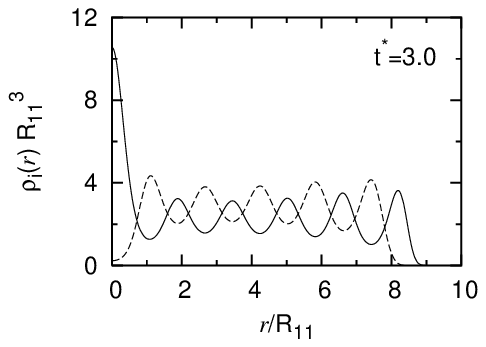}
\end{minipage}
\end{center}
\caption{Fluid A density profiles $\rho_i(r,t)$ (solid line: species 1,
dashed line:
species 2) for $N_1=N_2=6000$ particles. Initially
($t<0$) the fluid is at equilibrium confined in symmetric external potentials
with $E_2=E_1=10k_BT$ and ${\cal R}=8R_{11}$.
At $t=0$, $E_2$ is suddenly increased to
$E_2=20k_BT$, breaking the symmetry. The profiles are plotted for various
$t^*=k_BT \Gamma R_{11}^2 t$. Initially there was no signature of microphase
separation in the density profiles, but after breaking the symmetry, the fluid
profiles develop an `onion' structure.}
\label{fig:2}
\end{figure}

\begin{figure}[t]
\begin{center}
\begin{minipage}[t]{5.1cm}
\includegraphics[width=5cm]{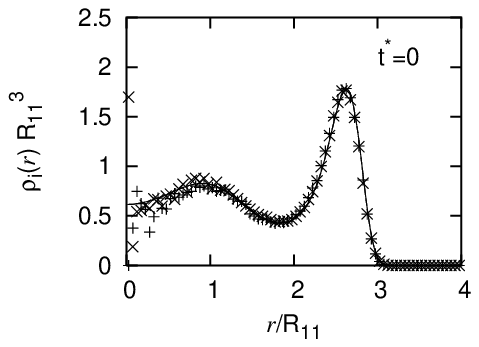}
\end{minipage}
\begin{minipage}[t]{5.1cm}
\includegraphics[width=5cm]{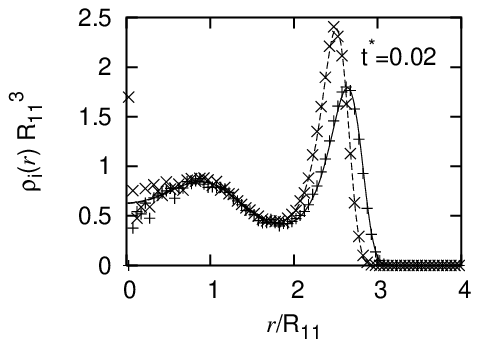}
\end{minipage}
\begin{minipage}[t]{5.1cm}
\includegraphics[width=5cm]{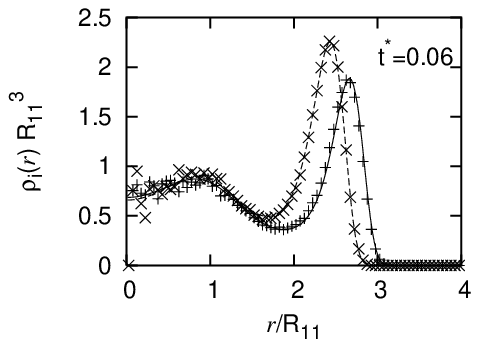}
\end{minipage}
\begin{minipage}[t]{5.1cm}
\includegraphics[width=5cm]{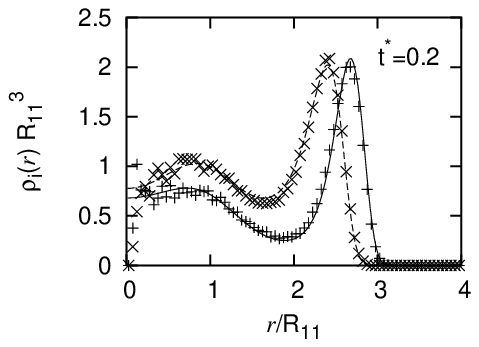}
\end{minipage}
\begin{minipage}[t]{5.1cm}
\includegraphics[width=5cm]{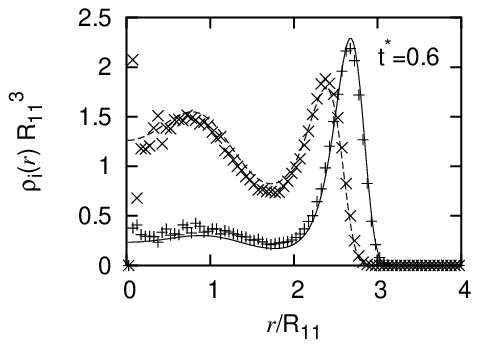}
\end{minipage}
\begin{minipage}[t]{5.1cm}
\includegraphics[width=5cm]{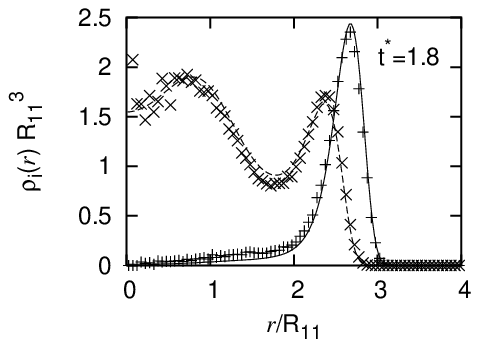}
\end{minipage}
\end{center}
\caption{As in Fig.\ \ref{fig:1}, but for fluid B with $N_1=N_2=100$.
The initial
density profiles are those for symmetric external potentials: $E_1=E_2=10k_BT$.
At $t=0$, $E_2 \rightarrow 20k_BT$. The subsequent profiles then show
the signature of phase separation.}
\label{fig:3}
\end{figure}

\begin{figure}[t]
\begin{center}
\begin{minipage}[t]{5.1cm}
\includegraphics[width=5cm]{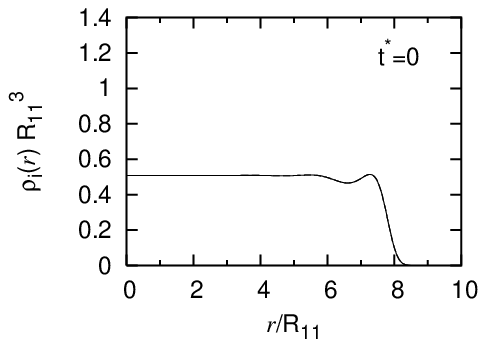}
\end{minipage}
\begin{minipage}[t]{5.1cm}
\includegraphics[width=5cm]{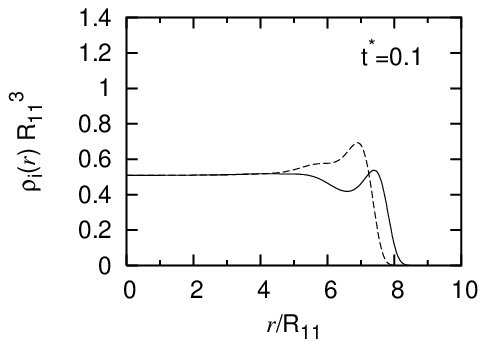}
\end{minipage}
\begin{minipage}[t]{5.1cm}
\includegraphics[width=5cm]{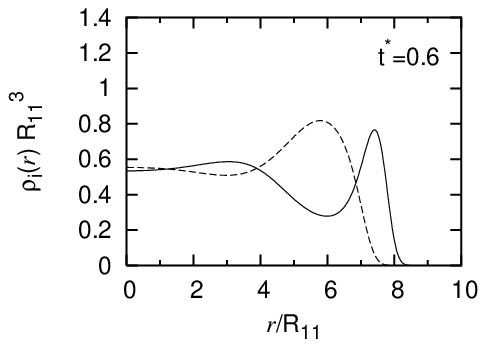}
\end{minipage}
\begin{minipage}[t]{5.1cm}
\includegraphics[width=5cm]{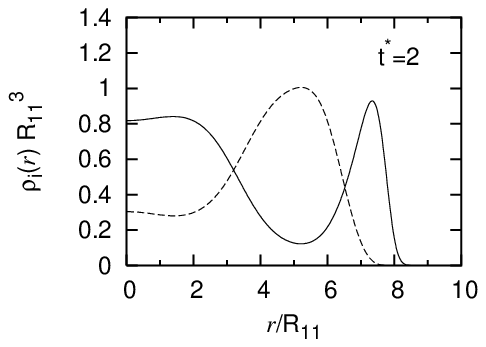}
\end{minipage}
\begin{minipage}[t]{5.1cm}
\includegraphics[width=5cm]{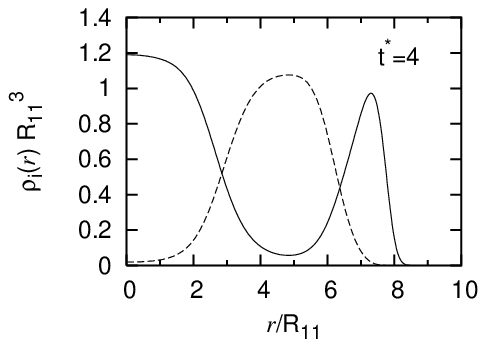}
\end{minipage}
\begin{minipage}[t]{5.1cm}
\includegraphics[width=5cm]{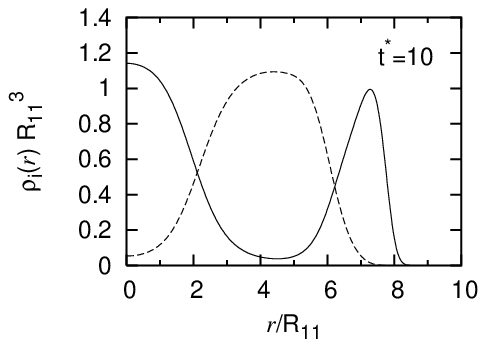}
\end{minipage}
\begin{minipage}[t]{5.1cm}
\includegraphics[width=5cm]{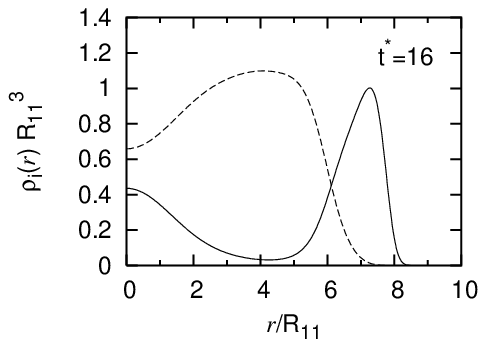}
\end{minipage}
\begin{minipage}[t]{5.1cm}
\includegraphics[width=5cm]{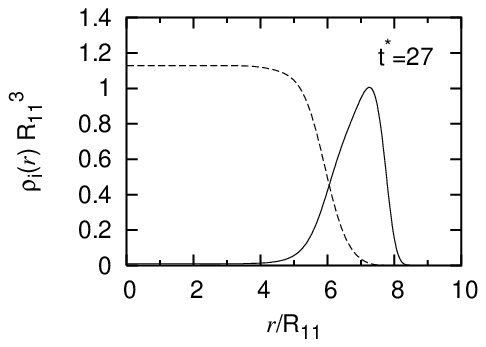}
\end{minipage}
\begin{minipage}[t]{5.1cm}
\includegraphics[width=4.5cm]{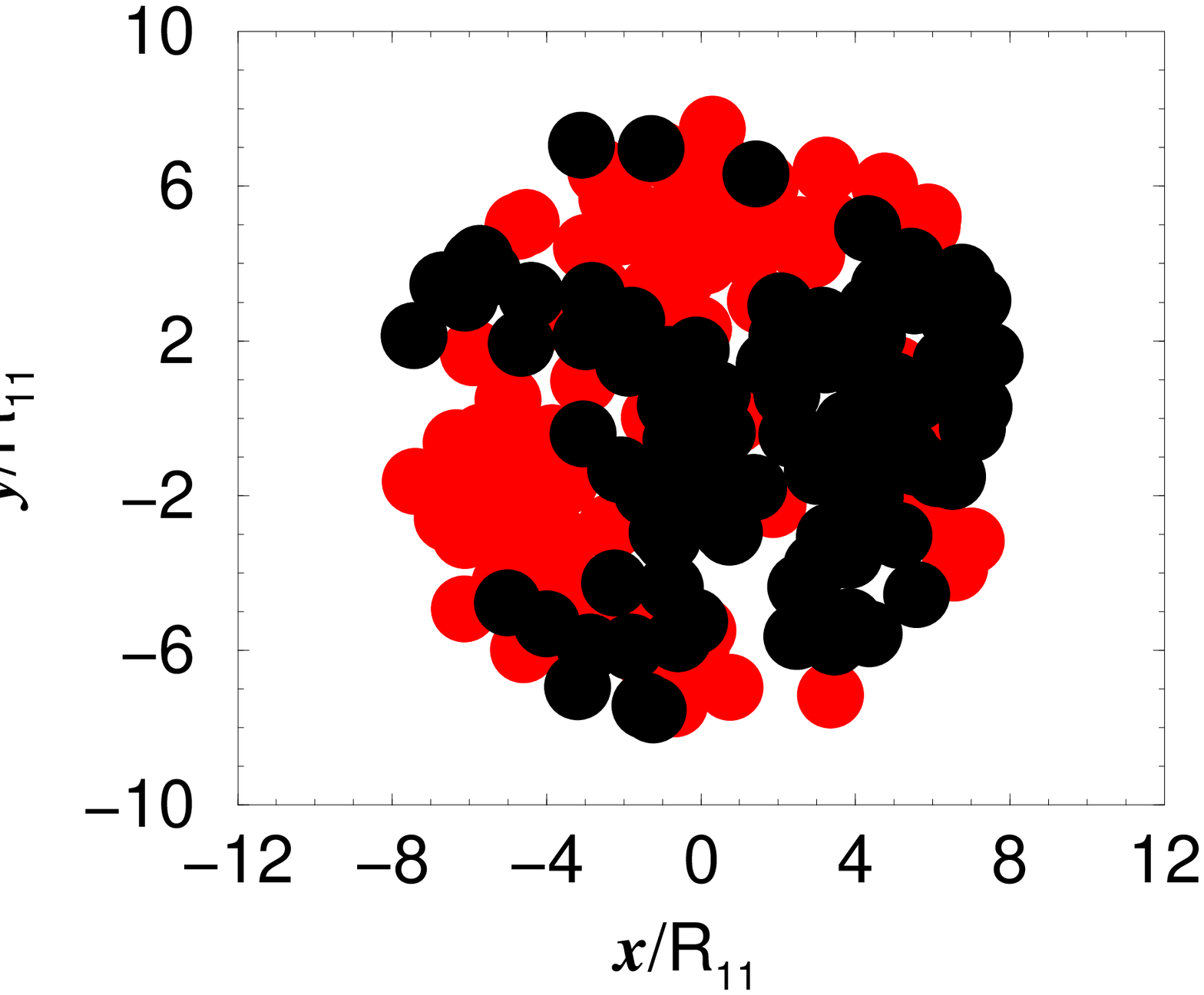}
\end{minipage}
\end{center}
\caption{As in Fig.\ \ref{fig:2}, but now for fluid B with $N_1=N_2=1000$
particles. Initially there is no signature of phase separation in the density
profiles, but after breaking the symmetry in the external potentials, there is.
Note the intermediate profiles exhibit {\em two} fluid-fluid
interfaces (rather than one). In the final frame we display a slice (particles
with $-0.5R_{11}<z<0.5R_{11}$) from an equilibrium configuration
$t<0$. Even though the one body
density profiles ($t=0$) show no signature of phase separation, this
configuration does.}
\label{fig:4}
\end{figure}

The present paper builds upon an earlier study \cite{Archer11},
in which we considered two different
{\em asymmetric} binary mixtures of
particles interacting via repulsive Gaussian pair potentials \cite{Likos}:
$u_{ij}(r) = \epsilon_{ij} \exp(-r^2/R_{ij}^2)$,
where $i,j=1,2$ label the two different species of particles,
$\epsilon_{ij}>0$ is a parameter that determines the strength
and $R_{ij}$ denotes the range of the interaction potential. For different
choices of these parameters one can obtain either a
mixture exhibiting bulk fluid--fluid (macro)phase separation \cite{paper1,
Archer1, Archer11}, or alternatively, a mixture exhibiting microphase separation
\cite{Archer6,Archer11}. We found that the DDFT used in conjunction with an
extremely simple approximate (RPA)
Helmholtz free energy functional was able to account accurately for the dynamics
of phase separation induced by reducing the radius of the confining spherical
cavity potential \cite{Archer11}.
In the present paper we consider similar situations but for {\em symmetric}
binary Gaussian core model (GCM) fluids, i.e.\ where $R_{11}=R_{22}$ and
$\epsilon_{11}=\epsilon_{22}$, but where $R_{12} \neq R_{11}$
or $\epsilon_{12} \neq \epsilon_{11}$.
The presence of this symmetry can have a
dramatic effect on the confined fluid one body density profiles in cases
where there are equal numbers of particles for each species.
As long as the confining cavity potentials are the same for both species of
particles, then the ensemble (or time) average one body density profiles of
the two species must be exactly the same. Symmetry dictates
this must always be the case, even when the fluid exhibits phase separation.
Here we show that when one breaks the symmetry of the
confining cavity potential, such that the cavity walls favour one of the
species, then the fluid one body density profiles can
change significantly and
show the influence of the phase separation -- see for example
the results in Figs.\ \ref{fig:1}--\ref{fig:4}.

We now describe further the model fluid and the DDFT: The GCM fluid
pair potentials have no hard core and the centres
of the particles can overlap completely. Such potentials arise when considers
the effective potential between the centres of mass of polymers in a good
solvent, see Ref.\ \cite{Likos} and references therein.
In this case $\epsilon_{ij} \sim
2k_BT$ and $R_{ij} \sim R_g$, the polymer radius of gyration. We consider
two symmetric binary mixtures of GCM particles. In both cases
$\epsilon_{11}=\epsilon_{22}=\epsilon_{12}=2k_BT$ and $R_{11}=R_{22}$. The first
mixture, which we denote ``fluid A'', is that with
$R_{12}=0.6R_{11}$. Fluid A exhibits microphase separation, and is
similar to the fluid described in Ref.\ \cite{Archer6}. This mixture
exhibits 1-2 ordering, i.e.\ where a particle of one species preferentially has
particles of the opposite species as nearest neighbours \cite{Archer6}.
The second mixture, which we denote ``fluid B'', is that with
$R_{12}=1.1R_{11}$. Fluid B exhibits macrophase or bulk phase separation. This
is driven by the fact that $R_{12}>R_{11}=R_{22}$ \cite{Archer1}. The bulk
critical point for this mixture is at a total density $\rho R_{11}^3=0.54$ and,
by symmetry, at concentration $x_1=x_2=0.5$.

When the GCM fluid density becomes sufficiently high, the following mean-field
Helmholtz free energy functional is rather accurate
\cite{Likos,paper1,Archer1}:
\begin{equation}
\fl
F[\{ \rho_i(\rr) \}]=F_{id}[\{ \rho_i(\rr) \}]+
\frac{1}{2}\sum_{i,j} \int \dr \int \dr'
\rho_i(\rr) \rho_j(\rr')u_{ij}(|\rr-\rr'|)
+\sum_i\int \dr V_i(\rr) \rho_i(\rr),
\nonumber
\end{equation}
where $\{\rho_i(\rr)\}$ are the fluid one body density profiles,
$V_i(\rr)$ is the external potential for particles of species $i$ and $F_{id}$
is the ideal gas contribution to the free energy \cite{Evans92}.

For a fluid of $N=N_1+N_2$ Brownian (colloidal) particles, one can approximate
the equations of motion using the following stochastic Langevin equations of
motion: $\Gamma_i^{-1} \dd \rr_n(t)/\dd t
=-\nabla_n U(\rr^N,t)+\GG_n(t)$,
where $\rr_n$ is the position of the $n^{th}$ particle, $\Gamma_i^{-1}$ is a
friction constant for particles of species $i$ (we assume
$\Gamma_1=\Gamma_2=\Gamma$), $U(\rr^N,t)$ is the potential
energy and $\GG_n(t)$ is a stochastic white noise term -- see Refs.\
\cite{Marconi:TarazonaJCP1999,Archer7,Archer11} for more details.
From such equations of motion, averaging over all
realisations of the stochastic noise, one can obtain the
following equations for the time evolution of the ensemble average
one body density profile
\cite{Marconi:TarazonaJCP1999,Marconi:TarazonaJPCM2000,Archer7,Archer11}:
\begin{equation}
\frac{\partial \rho_i(\rr,t)}{\partial t}
=\Gamma_i \nabla \cdot
\left[ \rho_i(\rr,t) \nabla
\left(\frac{\delta F[\{ \rho_i(\rr,t) \}]}{\delta
\rho_i(\rr,t)}\right)  \right],
\label{eq:mainres_multi}
\end{equation}
where the Helmholtz free energy functional $F[\{ \rho_i(\rr,t) \}]$ is given by
the equilibrium functional with the equilibrium density profiles
$\{\rho_i(\rr)\}$ replaced by the set of non-equilibrium profiles
$\{\rho_i(\rr,t)\}$.
We solve the DDFT (\ref{eq:mainres_multi}) for the binary GCM confined in
spherically symmetric external potentials of the form
\cite{joe:christos,Archer11}: $V_i(r) = E_i \left(r/{\cal R} \right)^{10}$,
where $r$ is the distance from the origin, $E_i$ is an energy scale and the
length-scale ${\cal R}$ (cavity radius)
is the same for both species of particles.

In Fig.\ \ref{fig:1} we display the results for $N_1=N_2=100$ particles of each
species of a mixture of fluid A. Initially ($t<0$) the external
potential parameters $E_2 \neq E_1$ and the fluid one body density profiles
exhibit 1-2 ordering (microphase separation) in the cavity. Then at $t=0$ the
external potentials are made symmetric: $E_1=E_2$. At equilibrium,
for such symmetric external potentials, the one body density profiles for each
species must be the same since $N_1=N_2$. We see in Fig.\ \ref{fig:1} that
the density profiles quickly evolve to give the same final profiles for each
species. We also display the results from BD simulations, which show extremely
good agreement with the results from the DDFT, giving confidence in the DDFT
approximation. In Fig.\ \ref{fig:2} we display DDFT results for a much larger
number of particles. The external potentials for $t<0$ are
symmetric and therefore the ensemble average density profiles at $t=0$ are the
same for both species. At $t=0$ the symmetry in the external potentials is
broken, such that
particles of species 1 are favoured by the walls of the cavity. The one body
density profiles then evolve towards equilibrium density profiles that display
the microphase ordering in the fluid \cite{Archer6}. Note that this does
not mean that there was no microphase separation for $t<0$ (the average total
density and composition has not changed); if one were to look at a `snapshot' of
the particle configurations at a time $t<0$ one would indeed see microphase
ordering. We illustrate this below for fluid B.

In Fig.\ \ref{fig:3} we display the results for $N_1=N_2=100$ particles for
fluid B. Initially ($t<0$) the external
potential parameters $E_2 = E_1$ and the fluid one body density profiles
are the same for each species. At $t=0$ the symmetry in the
external potentials is broken: $E_1 \neq E_2$, such that the cavity walls now
favour species 1. The equilibrium density profiles for the asymmetric
external potentials then show that the fluid does indeed exhibit the signature of
bulk phase separation. We also display the results from BD simulations,
which show extremely good agreement with the results from the DDFT.
In Fig.\ \ref{fig:4} we display the DDFT results for fluid B with
a larger number of particles. Once again the external potentials for $t<0$ are
symmetric and therefore the ensemble average density profiles at $t=0$ are
the same for both species.
At $t=0$ the symmetry is broken and the one body
density profiles then evolve towards ones that show the
fluid-fluid phase separation. Interestingly, this evolution to the final
equilibrium density profiles proceeds via an intermediate state where the fluid
is rich in species 1 at both the centre {\em and} close to the wall of the
cavity.
Then, as time proceeds, species 1 particles at the centre of the
cavity diffuse to the outside, since the intermediate state has two fluid-fluid
interfaces which has a higher free energy cost than the final equilibrium
configuration with only one fluid-fluid interface.
Despite the fact that the $t=0$ profiles show no sign of
phase separation, in the final frame of Fig.\ \ref{fig:4}
we see that in a particular configuration
of the particles that there is phase separation in the cavity -- i.e.\ the $t=0$
profiles in Fig.\ \ref{fig:4} are the result of ensemble averaging over strongly
asymmetric configurations. This implies that in this larger cavity
there are strong fluctuation
effects over the whole cavity, which are neglected in the present mean field
DFT treatment \cite{reguera04}. We therefore expect the DDFT results of Fig.\
\ref{fig:4} to be less reliable than those of Fig.\ \ref{fig:3}.

We conclude by emphasising that although the initial ($t=0$) one body density
profiles in Figs.\ \ref{fig:2} -- \ref{fig:4} show no obvious sign of phase
separation, this does not mean that there will be no signature of phase
separation in a {\em particular}
equilibrium configuration of the particles subject to symmetric external
potentials. However, in the ensemble average, the one body density profiles must
be the same and therefore show no signature of phase separation.
Such effects demonstrate an important aspect of the DDFT: this is a
theory for the {\em ensemble} average fluid one body density profiles
\cite{Archer8} and
therefore, as with equilibrium DFT, the fluid density profiles must exhibit the
symmetry of the confining external potentials.
%

{\bf Acknowledgements:}
I thank Bob Evans for a critical reading of the manuscript and
I gratefully acknowledge the support of EPSRC under grant number GR/S28631/01.

\end{document}